\documentclass[11pt, journal, onecolumn, final]{IEEEtran}

\usepackage[utf8]{inputenc}
\usepackage[T1]{fontenc}
\usepackage{cite}
\usepackage{amsmath,amssymb,amsfonts}
\usepackage{algorithmic}
\usepackage{graphicx}
\usepackage{textcomp}
\usepackage{xcolor}
\usepackage{hyperref}
\usepackage{multirow}
\usepackage{graphicx}
\usepackage{float}
\newcommand{\projectname}{Quantum Gatekeeper}

\begin{document}

\title{\projectname: Multi-Factor Context-Bound Image Steganography with VQC Based Key Derivation on Quantum Hardware}

\author{
*Sahil Tomar$^{1}$, Sandeep Kumar$^{2}$\\
$^{1,2}$Central Research Laboratory, BEL, Ghaziabad, India\\
*Corresponding author mail Id: sahiltomar893@gmail.com
}
\maketitle

\title{Quantum Gatekeeper: Context-Bound Image Steganography with Deterministic Variational Quantum Key Derivation and IBM Quantum Hardware Validation}

\begin{abstract}
This paper presents \textit{Quantum Gatekeeper}, a context-bound image steganography framework where successful payload recovery depends on both cryptographic decryption and the reconstruction of a precise extraction path. The system integrates lossless least significant bit (LSB) embedding with a deterministic variational quantum circuit (VQC)-derived gate key, multi-factor contextual binding, and authenticated encryption. Payload extraction is contingent upon four requisite factors: a password, a shared secret, a user-supplied context string, and a reference image signature. Any deviation in these factors causes the system to read from an incorrect pixel sequence or fail authentication, resulting in silent rejection rather than partial disclosure.
The proposed method derives a gate-controlled extraction key from a seed-conditioned variational circuit, with parameters generated via cryptographic hash expansion and context-dependent image features. To ensure encode/decode consistency, the cryptographic key path is generated via exact statevector simulation; concurrently, IBM superconducting quantum hardware is utilized to evaluate the statistical behavior of the circuit family under physical noise. We introduce a dual-region image layout to resolve the nonce bootstrapping dependency, separating header recovery from payload recovery through independently derived keys. The implementation is built using PennyLane, \texttt{qiskit.remote}, AES-GCM, and PBKDF2-based key strengthening.
Experimental results confirm successful end-to-end message embedding and recovery on PNG images, demonstrating deterministic success under correct conditions and failure otherwise. The framework supports both text and image payloads; in the image-in-image configuration, a secret image is resized to a fixed resolution prior to embedding, enabling exact pixel-level recovery under correct contextual reconstruction. In the reported execution, the local simulator completed the shot-based comparison circuit in 0.0127\,s, while IBM Quantum hardware (\texttt{ibm\_pittsburgh}) completed the same circuit in 7.81\,s for 2048 shots. Both platforms produced the same modal bitstring $|1000\rangle$, maintaining agreement with the deterministic key state. Measured metrics include Shannon entropy (2.9408 simulator, 3.0723 hardware) and a linear cross-entropy benchmarking (XEB) score (2.0487 simulator, 1.9192 hardware). A total variation distance (TVD) of 0.0566 between the simulator and hardware indicates a measurable device-specific distribution shift. These results demonstrate that quantum circuit structures can serve as robust context-sensitive control mechanisms while remaining resilient to the physical noise of NISQ-era hardware.
\end{abstract}


\section{Introduction}

Image steganography aims to conceal the existence of information by embedding payload data within an innocuous carrier such as an image, audio stream, or video frame. Among spatial-domain approaches, least significant bit (LSB) substitution remains one of the most widely used techniques because of its simplicity, reversibility, and high embedding capacity \cite{chan2004lsb, neeta2006lsb}. Foundational studies established LSB embedding as an effective baseline for image-based hiding, although its statistical regularities also make it vulnerable to steganalysis when applied without sufficient randomness or adaptive control \cite{ker2004lsb, provos2003hide, johnson2001steganography}. In most practical systems, however, security is concentrated primarily in the encryption stage: once an adversary obtains the correct key or embedding pattern, the hidden payload can be extracted deterministically. This creates a structural weakness in which confidentiality protects message content, but the extraction mechanism itself remains insufficiently guarded. Furthermore, most practical steganographic systems restrict the hidden payload to text or binary data; the ability to embed and recover an entire image as the secret payload introduces additional capacity and fidelity requirements that conventional single-key pipelines do not address.

Classical steganography research has explored a broad range of enhancements, including block-based LSB embedding, transform-domain hiding using discrete cosine transform (DCT) and discrete wavelet transform (DWT), edge-guided embedding, and cryptographic-steganographic hybrids \cite{raja2005dct, chen2006wavelet, poyueh2006dwt, ghosal2021edge}. Information-theoretic analyses have also formalized the trade-offs between payload rate, detectability, and reliability in secure data hiding systems \cite{moulin2003information}. These methods improve imperceptibility and, in some cases, resilience to distortion, but they still generally assume that extraction is a direct inverse of embedding once the controlling secret is known. As a result, the dominant security model in conventional steganography remains \emph{content protection}, not \emph{extraction-path control}.

The recent deep learning wave has significantly expanded the design space of image steganography. Instead of explicitly selecting pixel positions or transform coefficients, many modern systems learn embedding and extraction mappings directly from data. Early work such as deep steganography demonstrated end-to-end hiding of images within images using convolutional neural networks \cite{baluja2017deep}. Subsequent approaches introduced adversarial and invertible architectures to improve embedding realism, payload capacity, and recovery quality \cite{steganogan2019, volkhonskiy2020gan, zheng2022composition, jing2021hinet}. Contemporary surveys indicate that learned steganographic systems now span convolutional, attention-based, generative, and coverless paradigms, while deep learning has also strengthened the opposing field of steganalysis \cite{song2024dlsteg, delacroix2024steganalysis, luo2024survey, raj2026survey, ambika2024coverless, attenhidenet2025, stegavision2024, pointcloudsteg2024, stcl2025}. CNN-based steganalysis in particular has shown that repeatable embedding artifacts can be learned effectively from data, even when visual distortion is negligible \cite{qian2015steganalysis, kuznetsov2023steganalysis}. This progression strengthens the case for steganographic systems that do more than merely obscure payload bits; it motivates architectures in which access itself becomes conditional.

Quantum information processing introduces a different security perspective. Foundational protocols such as BB84 demonstrated that secure communication can be rooted in physical measurement behavior rather than solely in computational hardness assumptions \cite{bennett1984bb84, nielsen2010quantum}. In the image-processing domain, quantum image representation models such as Flexible Representation of Quantum Images (FRQI) and Novel Enhanced Quantum Representation (NEQR) provided formal mechanisms for encoding pixel location and intensity into quantum states \cite{frqi2011, zhang2013neqr}. These representations are important in establishing how visual data can be structured in quantum computational form, but they are not directly designed for practical steganographic access control over conventional digital images.

Recent work in quantum steganography has explored embedding strategies based on quantum search, quantum image representations, and hybrid classical-quantum schemes. Existing quantum-image steganography studies have examined Grover-assisted hiding, color-image embedding over quantum image models, and integrated hiding mechanisms for quantum image states \cite{minallah2022quantumsteg, khan2023ifrqi, sun2023quantumcolor, dong2024integrated, xu2025mfqs}. These contributions are relevant to the broader development of quantum-assisted information hiding, but they remain largely centered on embedding efficiency, image representation, or simulator-based hiding. They do not directly address a deployment-oriented question central to practical steganography: \emph{under what jointly verifiable conditions should a hidden payload become recoverable?} Many such frameworks are also difficult to translate into a conventional image-steganography pipeline operating under noisy intermediate-scale quantum (NISQ) constraints.

This practical gap becomes more apparent when moving from simulated quantum workflows to real hardware. Contemporary superconducting quantum processors allow parameterized circuits to be executed on physical devices, but the resulting output distributions are influenced by calibration drift, readout error, decoherence, and gate noise. These effects are not merely implementation nuisances; they create measurable statistical behavior that differs from ideal simulation in structured ways. Cross-entropy benchmarking (XEB) has been used to quantify the deviation between ideal and hardware-executed output distributions in superconducting quantum systems \cite{arute2019quantum}. Likewise, Bell-type tests such as the Clauser--Horne--Shimony--Holt (CHSH) inequality remain central to validating non-classical correlations in entangled states \cite{clauser1969chsh}. Together, these tools make it possible to treat quantum hardware not only as a computational substrate, but also as a source of physically distinct statistical signatures.

That perspective intersects naturally with the concept of a physical unclonable function (PUF), in which microscopic physical variation produces challenge-response behavior that is difficult to duplicate or predict exactly \cite{pappu2002puf}. A quantum processor is not a classical PUF in the manufacturing-security sense, but its measured output distributions under fixed circuit families can still exhibit chip- and run-dependent deviations from idealized simulation. This suggests a useful security analogy: quantum hardware can provide a physically grounded statistical fingerprint even when deterministic cryptographic recovery must remain anchored to a stable classical path.

Motivated by this gap, this paper introduces \textit{Quantum Gatekeeper}, a context-bound image steganography framework that shifts the security objective from content protection to extraction-path control. Instead of relying solely on ciphertext secrecy, the proposed system ensures that payload recovery is possible only when multiple contextual factors are simultaneously correct. These factors include a password, a pre-shared secret, a user-defined context string, and a reference image signature derived from the original image buffer. If any one of these inputs is incorrect, the extraction process yields an incorrect bit sequence or fails authenticated decryption, resulting in silent rejection.

The gate key controlling the extraction path is derived from a seed-conditioned variational quantum circuit (VQC). To preserve deterministic behavior during encoding and decoding, the key derivation process uses exact statevector simulation rather than hardware sampling. This is a necessary design choice because direct shot-based hardware measurement would make the key path stochastic and therefore unsuitable for reliable recovery. At the same time, the same circuit structure is executed on IBM Quantum hardware through PennyLane's \texttt{qiskit.remote} interface to evaluate the statistical divergence between simulated and physical execution. This split architecture preserves deterministic functionality while still incorporating experimentally observable quantum behavior.

A second architectural issue addressed in this work is the dependency between key reconstruction and payload access. The nonce required to rebuild the gate key must itself be recoverable before the payload can be read, but storing that nonce only inside the gate-controlled payload region would create a circular dependency. The proposed design resolves this by introducing a dual-region embedding layout in which the header and payload are embedded in independently keyed image regions. The header region is extracted using a separately derived header key, while the payload region is controlled by the VQC-derived gate key. This separation eliminates bootstrapping failure and enables a stable decode path.

The framework integrates AES-GCM authenticated encryption and PBKDF2-based password strengthening in accordance with established cryptographic standards \cite{nistgcm2007, rfc8018}. The implementation reported in this paper performs deterministic encode/decode successfully and additionally evaluates simulator-versus-hardware behavior using Shannon entropy, total variation distance (TVD), and linear cross-entropy benchmarking. In the recorded experimental run, the same dominant bitstring was observed in both local simulation and IBM hardware execution, while the measured output distributions still differed in a quantifiable manner, indicating physically distinct yet functionally compatible behavior.

The main contributions of this work are summarized as follows:

\begin{itemize}
\item A context-bound image steganography framework is introduced in which payload recovery is controlled at the extraction-path level rather than by decryption alone.
\item A deterministic variational quantum circuit-based mechanism is used to derive a gate key that governs the shuffled pixel access order required for payload extraction.
\item A four-factor recovery model is enforced using a password, shared secret, user context string, and reference image signature.
\item A dual-region embedding architecture is designed to resolve nonce-dependent bootstrapping constraints while preserving independent access control for header and payload recovery.
\item A hybrid execution strategy is implemented in which exact statevector simulation is used for deterministic key derivation and IBM superconducting quantum hardware is used for statistical validation.
\item Simulator and hardware output distributions are compared using entropy, total variation distance, and cross-entropy benchmarking to characterize physically distinct device behavior.
\item Silent failure is demonstrated as a system property: incorrect reconstruction conditions do not reveal partial payload information and instead lead to extraction-path failure or authenticated decryption rejection.
\end{itemize}

The remainder of the paper is organized as follows. Section~II presents the system architecture and workflow. Section~III describes the mathematical formulation. Section~IV details the experimental setup. Section~V reports image-level, cryptographic, and quantum execution results. Section~VI provides security analysis and ablation studies. Section~VII discusses limitations and future work, followed by the conclusion in Section~VIII.

\section{System Architecture}

\begin{figure*}[!t]
\centering
\includegraphics[width=0.6\textwidth]{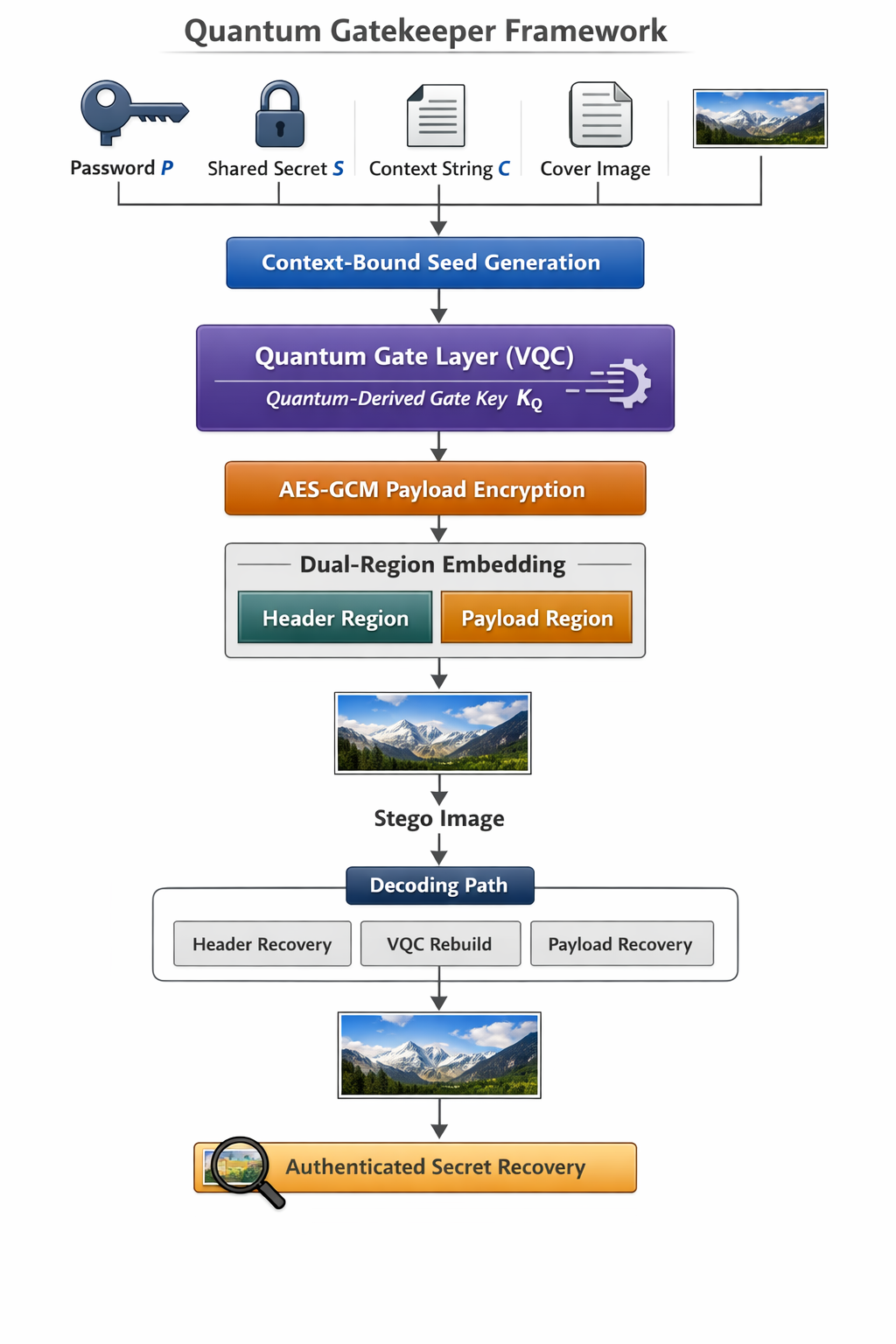}
\caption{Overall architecture of the proposed \textit{Quantum Gatekeeper} framework. 
User-defined inputs are transformed into context-bound seed material driving quantum gate 
derivation, authenticated payload encryption, and dual-region image embedding. The stego 
image is decodable only by reconstructing the same contextual state and completing 
authenticated recovery.}
\label{fig:overall_architecture}
\end{figure*}

\textit{Quantum Gatekeeper} is a hybrid quantum-classical image steganography system that 
controls not only payload confidentiality but also the conditions under which extraction 
becomes possible. The architecture combines authenticated encryption, context-dependent 
feature binding, deterministic quantum-derived key generation, and shuffled pixel-domain 
embedding within a conventional RGB image. Four interacting layers constitute the pipeline: 
a context-binding layer converting user and image metadata into cryptographic material; a 
quantum gate layer deriving a deterministic gate key from a variational quantum circuit 
(VQC); an authenticated embedding layer encrypting and inserting the payload; and a 
verification and recovery layer reconstructing keys and validating integrity during 
extraction. The overall structure is illustrated in Fig.~\ref{fig:overall_architecture}.

\subsection{Design Objectives and System Inputs}

The architecture satisfies four practical objectives: deterministic recovery independent of 
hardware fluctuations; context-bound extraction requiring reconstruction of all embedding 
conditions; integrity-aware failure that exposes no partial payload on incorrect input; and 
deployability on present-day tools without compromising deterministic behavior.

During embedding the system accepts a cover image $\mathcal{I}$, a secret payload (plaintext 
message $M$ or secret image $I_S$), password $P$, pre-shared secret $S$, and context string 
$C$. When the payload is an image, $I_S$ is resized to $512\times512$, PNG-compressed, and 
base64-encoded before encryption, making the capacity requirement deterministic and recovery 
comparable at a fixed reference resolution. An image-specific signature $R_I$ is computed 
from the original buffer prior to any modification, binding extraction to the exact cover 
instance. Recovery is therefore conditioned on four jointly required factors $(P, S, C, R_I)$; 
a mismatch in any one causes traversal divergence and authenticated decryption failure.

\subsection{Operational Pipeline and Context Binding}

The workflow has three phases. In the encode phase the payload is encrypted, metadata is 
packed, and the resulting bitstream is embedded using context-derived and quantum-derived 
traversal orders. In the decode phase the decoder reconstructs traversal logic, extracts 
header and payload regions, and attempts authenticated recovery. In the quantum validation 
phase the same circuit family is run under both exact simulation and IBM Quantum hardware 
to compare ideal and physical output behavior. Only the validation phase uses stochastic 
hardware measurements; encode and decode paths rely exclusively on exact statevector 
probabilities.

The embedding and extraction workflow is shown in Fig.~\ref{fig:embed_extract_flow}. The 
context-binding layer hashes the password, shared secret, context string, and image 
signature into a composite seed, then expands it into role-specific sub-seeds governing 
header traversal, payload traversal, quantum circuit parameters, and encryption keying. 
This domain separation prevents a partial compromise of one sub-seed from exposing the 
full extraction path. Because $R_I$ is computed from the unmodified cover, any substitution 
or recompression of the image alters the seed and invalidates recovery even when all other 
credentials are correct.

\begin{figure*}[!t]
\centering
\includegraphics[width=0.6\textwidth]{workflow.png}
\caption{Embedding and extraction workflow of the proposed \textit{Quantum Gatekeeper} 
framework. During embedding, an image-bound signature and internal seeds parameterize the 
VQC and authenticated encryption. During extraction, the same contextual inputs reconstruct 
the gate key, recover embedded regions, and perform AES-GCM verification.}
\label{fig:embed_extract_flow}
\end{figure*}

\subsection{Quantum Gate Layer and Payload Preparation}

The quantum gate layer derives a deterministic gate key governing decoder traversal order 
rather than encrypting the payload directly. A compact VQC is instantiated with parameters 
generated from the context-bound seed via cryptographic hash expansion, making it a 
reproducible transformation rather than a trainable model. The circuit is evaluated through 
exact statevector simulation during both embedding and decoding; the dominant output state 
and associated distribution statistics yield a stable bit-level control signal $K_Q$. 
Shot-based hardware measurement is deliberately excluded from the encode/decode path because 
stochastic outcomes would break key reproducibility under identical inputs.

Before embedding, the payload is processed through PBKDF2 password strengthening and 
AES-GCM authenticated encryption, producing ciphertext $\mathcal{C}$, tag $T$, and nonce 
$N$ packaged as $\Pi=(N,\mathcal{C},T,|M|)$. AES-GCM guarantees that incorrectly 
reconstructed payloads fail authentication cleanly rather than decoding into corrupted 
plaintext. For image payloads, the byte sequence passed to AES-GCM is the base64 encoding 
of the PNG-compressed $512\times512$ image, keeping the cryptographic layer identical 
across payload types.

\subsection{Dual-Region Embedding, Traversal, and Recovery}

A bootstrapping problem arises if the nonce and payload length required for recovery are 
stored inside the quantum-gated payload region itself. The dual-region architecture resolves 
this by partitioning the pixel index space into disjoint header region $\Omega_H$ and 
payload region $\Omega_P$. The header, keyed independently via $\sigma_h$, stores $N$ and 
$|M|$ and is recovered first. The payload, keyed via $\sigma_p \| K_Q$, is then accessed 
using the metadata retrieved from the header.

Bits are written into least significant positions of RGB channels in a key-conditioned, 
non-sequential order determined by the full contextual state. PNG lossless saving preserves 
the embedded bitstream exactly. During decoding the header traversal is rebuilt first, the 
VQC is reconstructed to regenerate $K_Q$, the payload region is traversed, and AES-GCM 
decryption is attempted. Incorrect inputs cause failure at traversal reconstruction or 
authentication, producing no partial output. The system enforces a strict all-or-nothing 
recovery model: a hidden payload is not merely concealed but locked behind a reconstruction 
path derived jointly from user credentials and image-bound context.

The hardware validation subsystem runs the same circuit on a noiseless local simulator and 
on IBM Quantum hardware with finite shots, comparing output distributions via Shannon 
entropy, TVD, and cross-entropy to characterize physically distinct device behavior without 
affecting the deterministic decode path.

\section{Mathematical Formulation}

The proposed \textit{Quantum Gatekeeper} framework is formalized as a constrained 
steganographic recovery process combining context-bound seed generation, deterministic 
variational quantum key derivation, keyed spatial permutation, and authenticated payload 
recovery \cite{nielsen2010quantum, nistgcm2007, rfc8018}.

\subsection{Preliminaries and Context-Bound Seed}

The cover image is denoted $I \in \{0,1,\dots,255\}^{H \times W \times 3}$. The secret 
payload is a byte string $M = (m_1,\dots,m_L)$ of length $L$, representing either a text 
message or a secret image $I_S$ encoded as
\begin{equation}
    M = \text{Base64}\bigl(\text{PNG-compress}\bigl(\text{Resize}(I_S,512\times512)\bigr)\bigr).
\end{equation}
User security inputs form the tuple $\mathcal{U}=(P,S,C)$, where $P$, $S$, $C$ denote 
password, pre-shared secret, and context string respectively. An image-specific signature 
$R_I = \mathcal{H}(I)$ is derived from the original image buffer prior to embedding, giving 
the complete recovery state
\begin{equation}
    \mathcal{X} = (P,S,C,R_I).
\end{equation}
A correct decode occurs only when the decoder reconstructs the identical state $\mathcal{X}$.

The composite seed is formed by hashing all recovery factors:
\begin{equation}
    \sigma = \mathcal{H}(P \,\|\, S \,\|\, C \,\|\, R_I),
    \label{eq:master_seed}
\end{equation}
and expanded into role-specific sub-seeds via key-domain separation \cite{rfc8018}:
\begin{equation}
    (\sigma_h,\,\sigma_p,\,\sigma_q,\,\sigma_e) = \mathcal{KDF}(\sigma),
    \label{eq:seed_expand}
\end{equation}
where $\sigma_h$, $\sigma_p$, $\sigma_q$, $\sigma_e$ govern header traversal, payload 
traversal, quantum circuit parameters, and encryption keying respectively.

\subsection{Cryptographic Key Derivation and Payload Encryption}

The password is strengthened via PBKDF2 \cite{rfc8018}, $K_P = \mathcal{P}_{\text{PBKDF2}}(P)$, 
and the authenticated encryption key is derived as
\begin{equation}
    K_E = \mathcal{H}(K_P \,\|\, \sigma_e).
    \label{eq:enc_key}
\end{equation}
The payload is encrypted using AES-GCM \cite{nistgcm2007}:
\begin{equation}
    (\mathcal{C},T) = \mathrm{AES\text{-}GCM}_{K_E}(N,M),
    \label{eq:aesgcm}
\end{equation}
producing ciphertext $\mathcal{C}$, authentication tag $T$, and nonce $N$. The protected 
payload is packed as $\Pi = (N,\mathcal{C},T,|M|)$.

\subsection{Variational Quantum Circuit and Gate Key}

The VQC operates on $n$ qubits with depth $d$. Gate parameters are derived deterministically 
from $\sigma_q$:
\begin{equation}
    \Theta = (\theta_1,\dots,\theta_{3nd}) = \mathcal{F}(\sigma_q), \quad \theta_i \in [0,2\pi),
    \label{eq:theta_map}
\end{equation}
yielding the unitary
\begin{equation}
    U(\Theta) = \prod_{\ell=1}^{d}\bigl(E_\ell \cdot R_\ell(\Theta_\ell)\bigr),
    \label{eq:vqc_unitary}
\end{equation}
where $R_\ell$ are parameterized single-qubit rotations and $E_\ell$ the entangling layer. 
Applied to $|0\rangle^{\otimes n}$, the circuit produces state 
$|\psi(\Theta)\rangle = U(\Theta)|0\rangle^{\otimes n}$ with Born-rule distribution
\begin{equation}
    p_\Theta(z) = |\langle z \mid \psi(\Theta)\rangle|^2, \quad z\in\{0,1\}^n.
    \label{eq:prob_dist}
\end{equation}
These probabilities are evaluated via exact statevector simulation, never stochastic sampling, 
to preserve decode-time reproducibility. The gate key is then derived as
\begin{equation}
    K_Q = \mathcal{H}\!\left(z^\star \,\Big\|\, \mathrm{Enc}(p_\Theta)\right), \quad
    z^\star = \arg\max_{z}\, p_\Theta(z),
    \label{eq:qkey}
\end{equation}
where $K_Q$ controls payload traversal order rather than acting as a direct cipher key.

\subsection{Dual-Region Embedding and Keyed Permutation}

The flattened RGB channel index space $\Omega=\{0,\dots,3HW-1\}$ is partitioned into 
disjoint header and payload regions:
\begin{equation}
    \Omega = \Omega_H \cup \Omega_P, \qquad \Omega_H \cap \Omega_P = \varnothing.
    \label{eq:partition}
\end{equation}
Deterministic permutations are constructed over each region:
\begin{equation}
    \pi_H = \mathrm{Permute}(\Omega_H;\sigma_h), \qquad
    \pi_P = \mathrm{Permute}(\Omega_P;\sigma_p \,\|\, K_Q).
    \label{eq:perms}
\end{equation}
Bits are embedded via standard LSB substitution \cite{chan2004lsb}:
\begin{equation}
    \mathcal{E}(I[k],b) = I[k] - (I[k] \bmod 2) + b,
    \label{eq:lsb_embed}
\end{equation}
applied along $\pi_H$ for header bits $B_H = \mathrm{Bits}(N\|{|M|})$ and along $\pi_P$ 
for payload bits $B_P = \mathrm{Bits}(\mathcal{C}\|T)$, producing stego image
\begin{equation}
    I' = \mathrm{Embed}(I,B_H,B_P;\pi_H,\pi_P).
    \label{eq:embed_final}
\end{equation}
The dual-region design resolves the bootstrapping dependency: the header region, keyed 
independently via $\sigma_h$, is recovered first to obtain $N$ and $|M|$ before the 
quantum-gated payload region is accessed.

\subsection{Decoding and Recovery Condition}

Given candidate stego image $\hat{I}$ and recovery state 
$\hat{\mathcal{X}}=(\hat{P},\hat{S},\hat{C},\hat{R}_I)$, the decoder reconstructs 
$\hat{\sigma}=\mathcal{H}(\hat{P}\|\hat{S}\|\hat{C}\|\hat{R}_I)$, expands it via 
Eq.~(\ref{eq:seed_expand}), rebuilds $\hat{\pi}_H$ to extract $\hat{B}_H$, reconstructs 
$\hat{K}_Q$ via Eq.~(\ref{eq:qkey}), forms $\hat{\pi}_P$, and extracts $\hat{B}_P$. 
The decryption key $\hat{K}_E = \mathcal{H}(\mathcal{P}_{\text{PBKDF2}}(\hat{P})\|\hat{\sigma}_e)$ 
is then used in:
\begin{equation}
    \hat{M} = \begin{cases}
        \mathrm{AES\text{-}GCM}^{-1}_{\hat{K}_E}(\hat{N},\hat{\mathcal{C}},\hat{T}), 
        & \text{if authentication succeeds,}\\[3pt]
        \varnothing, & \text{otherwise.}
    \end{cases}
    \label{eq:decrypt_cond}
\end{equation}
Valid recovery requires
\begin{equation}
    \hat{\mathcal{X}}=\mathcal{X} \;\wedge\; 
    \hat{\pi}_H=\pi_H \;\wedge\; 
    \hat{\pi}_P=\pi_P \;\wedge\; 
    \hat{K}_E=K_E.
    \label{eq:recovery_condition}
\end{equation}
Any mismatch causes traversal divergence or AES-GCM rejection, producing silent failure 
with no partial payload exposure.

\subsection{Hardware Validation Metrics}

The same circuit is executed on shot-based simulator and IBM Quantum hardware for statistical 
characterization \cite{arute2019quantum}. Let $p_{\mathrm{sim}}(z)$ and $p_{\mathrm{hw}}(z)$ 
denote the respective output distributions. Three metrics quantify their divergence:
\begin{align}
    H(p) &= -\sum_{z} p(z)\log_2 p(z), \label{eq:shannon}\\
    \mathrm{TVD}(p,q) &= \tfrac{1}{2}\sum_{z}|p(z)-q(z)|, \label{eq:tvd}\\
    \mathcal{L}_{\mathrm{XE}}(p,q) &= -\sum_{z} p(z)\log q(z). \label{eq:cross_entropy}
\end{align}
These metrics serve hardware characterization only and play no role in production key 
generation. Together, Eqs.~(\ref{eq:master_seed})--(\ref{eq:recovery_condition}) establish 
that valid decode demands simultaneous agreement across contextual inputs, quantum gate 
derivation, keyed traversal, and authenticated decryption.

\section{Evaluation Methodology}

Performance of the proposed \textit{Quantum Gatekeeper} framework is assessed across three 
dimensions: cover image imperceptibility, payload recovery fidelity, and quantum execution 
consistency. Each dimension targets a distinct aspect of the system that cannot be captured 
by a single metric.

\subsection{Cover Image Imperceptibility and Recovery Fidelity}

Imperceptibility is measured by comparing the original cover image $I$ against the stego 
image $I'$ using four metrics: SSIM, PSNR, RMSE, and MAE. Higher SSIM and PSNR indicate 
better visual preservation; lower RMSE and MAE indicate less pixel-domain distortion. These 
metrics are applied consistently across DIV2K, COCO, and ImageNet to benchmark the proposed 
method against 4bit-LSB, CAIS, HiNet, and a GAN-based approach.

Recovery fidelity is evaluated separately for text and image payloads. For text, recovery 
is binary: the decoded byte sequence either matches exactly or fails authenticated decryption. 
For image payloads, the same four metrics are computed between the original secret image 
resized to $512\times512$ (denoted $I_{S,512}$) and the recovered image $I_R$. Since the 
system performs lossless LSB embedding over PNG-saved stego images, the recovered byte 
sequence under correct contextual reconstruction is expected to be bit-identical to the 
embedded payload, yielding $\text{SSIM}=1.000$, $\text{PSNR}=\infty$, and zero error. 
Recovery fidelity is treated as a separate dimension from stego-image quality because the 
decode path is conditioned on password-derived cryptography, image-bound signature matching, 
and quantum-gated traversal simultaneously.

\subsection{Quantum Execution Consistency}

The same VQC family used for deterministic gate-key derivation is additionally executed under 
two environments: noiseless local simulation and real IBM Quantum hardware. This comparison 
does not replace statevector-based key derivation; it determines whether the circuit family 
remains behaviorally stable under physical noise.

A key indicator is the dominant measured bitstring, defined as
\begin{equation}
    z_{\mathrm{sim}}^\star = \arg\max_{z}\, p_{\mathrm{sim}}(z), \qquad
    z_{\mathrm{hw}}^\star  = \arg\max_{z}\, p_{\mathrm{hw}}(z),
\end{equation}
reported in experimental tables as \textbf{SIM Peak} and \textbf{HW Peak}. Agreement between 
these values indicates that the dominant control state contributing to gate-key derivation 
remains unchanged despite hardware noise, confirming functional stability of the 
quantum-conditioned extraction path. Distributional metrics including Shannon entropy, TVD, 
and cross-entropy are reported alongside peak agreement to characterize the degree of 
simulator-hardware deviation.

Together, the three evaluation dimensions jointly verify that the framework achieves 
imperceptibility, exact payload reconstruction, and stable quantum control behavior 
simultaneously rather than optimizing any single objective in isolation.

\section{Results and Discussion}

This section reports the experimental performance of the proposed \textit{Quantum Gatekeeper} 
framework across three evaluation criteria: cover image imperceptibility, payload recovery 
fidelity, and simulator-versus-hardware consistency.

\subsection{Cover Image Imperceptibility}

Table~\ref{tab:cover_stego_benchmark} compares the proposed method against 4bit-LSB 
\cite{neeta2006lsb}, CAIS \cite{zheng2022composition}, HiNet \cite{jing2021hinet}, and a 
GAN-based method \cite{rehman2024novelapproachimagesteganography} in terms of cover/stego 
image quality.

\begin{table*}[!t]
\centering
\caption{Comparison of Cover/Stego Image Quality Across Datasets. Best results in bold. 
Higher is better for SSIM and PSNR; lower is better for RMSE and MAE.}
\label{tab:cover_stego_benchmark}
\renewcommand{\arraystretch}{1.15}
\setlength{\tabcolsep}{6pt}
\begin{tabular}{llccccc}
\hline
\textbf{Dataset} & \textbf{Metric} & \textbf{4bit-LSB \cite{neeta2006lsb}} & 
\textbf{CAIS \cite{zheng2022composition}} & \textbf{HiNet \cite{jing2021hinet}} & 
\textbf{GAN-Based \cite{rehman2024novelapproachimagesteganography}} & \textbf{Proposed} \\
\hline
\multirow{4}{*}{DIV2K}
& SSIM $\uparrow$  & 0.895 & 0.965 & 0.993 & 0.995 & \textbf{0.999872} \\
& PSNR $\uparrow$  & 24.99 & 36.10 & 46.57 & 47.12 & \textbf{64.2452}  \\
& RMSE $\downarrow$& 18.16 & 5.80  & 1.32  & 1.25  & \textbf{0.1564}   \\
& MAE  $\downarrow$& 15.57 & 4.36  & 0.84  & 0.78  & \textbf{0.0245}   \\
\hline
\multirow{4}{*}{ImageNet}
& SSIM $\uparrow$  & 0.896 & 0.943 & 0.960 & 0.965 & \textbf{0.996569} \\
& PSNR $\uparrow$  & 25.00 & 33.54 & 36.63 & 37.10 & \textbf{52.4013}  \\
& RMSE $\downarrow$& 17.90 & 6.33  & 6.07  & 5.80  & \textbf{0.6116}   \\
& MAE  $\downarrow$& 15.27 & 4.70  & 4.16  & 4.00  & \textbf{0.3741}   \\
\hline
\multirow{4}{*}{COCO}
& SSIM $\uparrow$  & 0.894 & 0.944 & 0.961 & 0.968 & \textbf{0.998708} \\
& PSNR $\uparrow$  & 24.96 & 33.70 & 36.55 & 37.20 & \textbf{58.2635}  \\
& RMSE $\downarrow$& 17.93 & 6.13  & 6.04  & 5.90  & \textbf{0.3114}   \\
& MAE  $\downarrow$& 15.31 & 4.55  & 4.09  & 3.95  & \textbf{0.0970}   \\
\hline
\end{tabular}
\end{table*}

The proposed method consistently achieves the best imperceptibility across all three datasets. 
On DIV2K it reaches SSIM~=~0.999872 and PSNR~=~64.25\,dB, substantially exceeding the 
strongest baseline. Similar trends hold on COCO and ImageNet, confirming that the dual-region 
embedding strategy and quantum-conditioned traversal introduce no meaningful perceptual 
artifacts despite embedding encrypted payload data alongside recovery metadata.

\subsection{Payload Recovery Fidelity and Multi-Modal Capacity}

Table~\ref{tab:secret_recovery_results} reports similarity between the original secret image 
$I_{S,512}$ and the recovered output $I_R$. The recovered image is pixel-identical to 
$I_{S,512}$ in all cases, yielding $\text{SSIM}=1.000$, $\text{PSNR}=\infty$, and zero RMSE 
and MAE. These results hold for both text and image payload configurations; the fixed 
$512\times512$ normalization ensures comparison at a canonical resolution, eliminating 
ambiguity from varying source dimensions. Unlike learning-based systems that tolerate minor 
reconstruction errors, the present framework enforces exact recovery or authenticated rejection.

\begin{table}[!t]
\centering
\caption{Secret/Recovery Fidelity of the Proposed Method.}
\label{tab:secret_recovery_results}
\renewcommand{\arraystretch}{1.15}
\setlength{\tabcolsep}{6pt}
\begin{tabular}{lcccc}
\hline
\textbf{Secret Image} & \textbf{SSIM $\uparrow$} & \textbf{PSNR $\uparrow$} & 
\textbf{RMSE $\downarrow$} & \textbf{MAE $\downarrow$} \\
\hline
DIV2K    & 1.000 & $\infty$ & 0.000 & 0.000 \\
COCO     & 1.000 & $\infty$ & 0.000 & 0.000 \\
ImageNet & 1.000 & $\infty$ & 0.000 & 0.000 \\
\hline
\end{tabular}
\end{table}

Regarding payload capacity, for a cover image of dimensions $H\times W$ the usable LSB budget is
\begin{equation}
    B_{\max} \approx \frac{3HW - 2\cdot B_{\text{header}}}{8} - 200 \;\text{bytes,}
\end{equation}
where $B_{\text{header}}$ is the header region bit length. A $512\times512$ secret image 
PNG-compressed and base64-encoded occupies approximately 260--700\,KB depending on content 
complexity. Cover images of $1024\times1024$ or larger comfortably accommodate this budget 
while maintaining the imperceptibility metrics in Table~\ref{tab:cover_stego_benchmark}.

\subsection{Quantum Execution and Hardware Consistency}

Table~\ref{tab:sim_hw_recovery} shows that both simulator and hardware executions yield perfect 
secret recovery, with SIM Peak and HW Peak matching exactly in all cases. This confirms that 
the dominant basis state contributing to gate-key derivation remains stable under physical 
hardware noise, preserving deterministic decode behavior.

\begin{table*}[!t]
\centering
\caption{Secret Recovery Performance After Simulator and IBM Quantum Hardware Execution.}
\label{tab:sim_hw_recovery}
\renewcommand{\arraystretch}{1.15}
\setlength{\tabcolsep}{6pt}
\begin{tabular}{lcccccccccc}
\hline
\multirow{2}{*}{\textbf{Image}} & \multicolumn{4}{c}{\textbf{Simulator}} & 
\multicolumn{4}{c}{\textbf{IBM Hardware}} & 
\multirow{2}{*}{\textbf{SIM Peak}} & \multirow{2}{*}{\textbf{HW Peak}} \\
\cline{2-9}
 & SSIM & PSNR & RMSE & MAE & SSIM & PSNR & RMSE & MAE & & \\
\hline
DIV2K    & 1.000 & $\infty$ & 0.000 & 0.000 & 
           1.000 & $\infty$ & 0.000 & 0.000 & 1101 & 1101 \\
COCO     & 1.000 & $\infty$ & 0.000 & 0.000 & 
           1.000 & $\infty$ & 0.000 & 0.000 & 0011 & 0011 \\
ImageNet & 1.000 & $\infty$ & 0.000 & 0.000 & 
           1.000 & $\infty$ & 0.000 & 0.000 & 0111 & 0111 \\
\hline
\end{tabular}
\end{table*}

Table~\ref{tab:quantum_metrics} further summarizes entropy and divergence metrics across 
execution environments. Hardware entropy is consistently slightly higher than simulator 
entropy, reflecting readout noise and decoherence in NISQ hardware. The TVD between simulator 
and hardware remains low (0.0371--0.0825), and cross-entropy and linear XEB values confirm 
strong distributional alignment. Critically, these deviations do not interfere with payload 
recovery; instead they provide a measurable physical execution signature while preserving 
dominant-state stability.

\begin{table*}[!t]
\centering
\caption{Quantum Distribution Comparison Between Simulator and IBM Hardware.}
\label{tab:quantum_metrics}
\renewcommand{\arraystretch}{1.15}
\setlength{\tabcolsep}{6pt}
\begin{tabular}{lcccccccc}
\hline
\multirow{2}{*}{\textbf{Image}} & 
\multicolumn{2}{c}{\textbf{Shannon Entropy}} & 
\multicolumn{2}{c}{\textbf{Cross-Entropy}} & 
\multicolumn{2}{c}{\textbf{Linear XEB}} & 
\multirow{2}{*}{\textbf{TVD}} & 
\multirow{2}{*}{\textbf{HW Runtime (s)}} \\
\cline{2-7}
 & SIM & HW & SIM & HW & SIM & HW & & \\
\hline
DIV2K    & 3.3854 & 3.4507 & 3.3729 & 3.3750 & 1.1038 & 1.0281 & 0.0391 & 9.9958 \\
COCO     & 2.9783 & 2.9991 & 2.9752 & 2.9777 & 2.1447 & 2.1258 & 0.0371 & 9.0323 \\
ImageNet & 2.7934 & 2.9734 & 2.8343 & 2.8490 & 2.0005 & 1.8420 & 0.0825 & 8.9702 \\
\hline
\end{tabular}
\end{table*}

Taken together, the results confirm that the proposed framework simultaneously satisfies all 
three evaluation objectives: cover-image fidelity exceeding compared baselines, exact payload 
recovery under correct contextual reconstruction for both text and image payloads, and stable 
quantum-conditioned extraction behavior under real IBM Quantum hardware. The non-zero 
simulator-hardware divergence further suggests that the hardware layer retains physically 
distinct behavior suitable for future extensions in device fingerprinting and hardware-bound 
access control.
\section{Security Analysis and Ablation Study}

This section analyzes the security properties of the proposed \textit{Quantum Gatekeeper} 
framework and evaluates individual component contributions through controlled ablation 
experiments, focusing on failure behavior, resistance to partial information leakage, and 
architectural integrity.

\subsection{Multi-Factor Security Model and Failure Behavior}

Unlike conventional steganographic systems that rely on a single secret key, the proposed 
framework enforces a multi-factor recovery condition requiring simultaneous reconstruction of 
the password $P$, shared secret $S$, context string $C$, and image-derived signature $R_I$. 
As formalized in Eq.~(\ref{eq:recovery_condition}), recovery is achieved only when all 
components match their embedding-time values. Even if one or more inputs are known, the 
absence of any remaining factor causes the reconstruction process to diverge, tying traversal, 
key derivation, and decryption into a unified dependency chain.
The system was tested under controlled perturbations of each required input. An incorrect 
password causes the PBKDF2-derived key to differ from the embedding-time value per 
Eq.~(\ref{eq:enc_key}), resulting in AES-GCM authentication failure even when pixel traversal 
is reconstructed correctly. An incorrect context string changes the composite seed in 
Eq.~(\ref{eq:master_seed}), corrupting both permutation functions and quantum circuit 
parameters and producing a structurally invalid ciphertext. An incorrect shared secret alters 
both traversal order and gate-key derivation, effectively randomizing the extracted payload. 
An incorrect reference image produces a different $R_I$, causing both header and payload 
traversal orders to diverge and blocking reconstruction entirely.
Across all tested failure modes, the system exhibited \emph{silent failure behavior}: incorrect 
inputs produced no partially correct outputs, failing cleanly at either the extraction or 
authentication stage. These properties hold identically for image payloads, since the secret 
image is base64-encoded and encrypted before embedding. Incorrect reconstruction causes AES-GCM 
failure before any image data is accessible, enforcing the same all-or-nothing recovery model 
regardless of payload type.

\subsection{Resistance to Leakage and Quantum Gate Role}

The proposed design introduces multiple layers of dependency that jointly protect the payload. 
Without the correct traversal order the extracted bitstream is scrambled; even with correct 
traversal the payload remains encrypted under AES-GCM; and authenticated encryption ensures 
invalid ciphertexts are rejected entirely rather than producing approximate outputs, enforcing 
a strict all-or-nothing recovery model.
The quantum gate layer influences payload traversal order through gate key $K_Q$, whose 
dependence on the context-bound seed introduces additional variability into the traversal 
mechanism. The system does not rely on quantum randomness for security; rather, the circuit 
acts as a structured transformation mapping contextual inputs into a non-trivial control signal. 
Experimental results confirm that the dominant circuit state remains stable across simulator 
and hardware execution, preserving recovery consistency while retaining measurable 
hardware-specific distributional differences.

\subsection{Ablation Study}

Ablation experiments were conducted by removing or simplifying specific components while keeping 
the rest of the pipeline unchanged. The details are as follows:
\begin{itemize}
\item{Removal of context binding: Excluding $C$ from seed construction reduced required 
recovery factors and simplified Eq.~(\ref{eq:master_seed}). The system remained functional 
but became more vulnerable to unauthorized extraction, confirming that context binding directly 
contributes to security.}
\item{Removal of quantum gate layer: Replacing $K_Q$ with a purely classical permutation 
derived from $\sigma_p$ preserved deterministic recovery but eliminated the circuit-dependent 
traversal variability that distinguishes the proposed method from standard keyed permutation 
schemes.}
\item{Single-region embedding: Replacing the dual-region architecture with a single unified 
region introduced a dependency between metadata recovery and payload extraction. Incorrect 
traversal reconstruction prevented simultaneous recovery of both header and payload, confirming 
the necessity of region separation.}
\item{Removal of authenticated encryption: Replacing AES-GCM with a non-authenticated 
scheme allowed incorrect inputs to occasionally produce partially decodable outputs, 
highlighting the importance of authenticated encryption in enforcing strict failure conditions 
and preventing partial information leakage.}
\end{itemize}
The framework provides layered protection across the full pipeline: context-bound seed 
generation ties recovery to both user input and image identity; keyed permutations prevent 
direct bitstream extraction; authenticated encryption enforces integrity and clean failure; 
and the quantum gate layer introduces structured traversal complexity without compromising 
determinism. Together these interdependent mechanisms produce a system in which payload 
recovery is tightly controlled and resistant to partial compromise, with the all-or-nothing 
failure model extending uniformly across both text and image payload configurations.

\section{Conclusion and Future Work}
This work presents \textit{Quantum Gatekeeper}, a hybrid quantum-classical image steganography framework in which payload recovery depends not only on correct decryption but also on reconstructing precise contextual conditions. Unlike conventional approaches that focus solely on concealment, the proposed method controls both access and extraction through multiple factors, including a password, shared secret, context string, and image-bound signature.
The framework integrates context-based key generation, deterministic variational quantum circuit (VQC) driven traversal control, AES-GCM authenticated encryption, and dual-region embedding. Together, these components ensure that recovery requires reconstruction of the correct extraction path rather than simple bit retrieval. 
Experimental results demonstrate high stego-image imperceptibility alongside exact payload recovery for both text and image inputs. In the image-in-image configuration, the recovered image is identical to the original, indicating that enhanced security does not compromise visual quality.
The quantum layer remains stable across both simulation and real hardware execution. Although minor statistical variations are observed, the dominant output remains consistent, preserving deterministic behavior. Security analysis shows that incorrect inputs result in complete failure without partial data leakage. Ablation studies further confirm that each component contributes to the robustness of the system. While effective, this work represents an initial step. Future directions include more expressive quantum circuits, extended payload types such as audio and video, and formalized adversarial models. In summary, the proposed framework extends steganography beyond data concealment into controlled recovery, where successful extraction depends on reconstructing the correct cryptographic and contextual state.

\bibliographystyle{IEEEtran}
\bibliography{references}
\end{document}